\documentclass[a4paper]{article}
\usepackage{ASVspoof}
\usepackage{epsfig,amssymb,amsmath}
\usepackage[hidelinks]{hyperref}
\usepackage{graphicx}
\usepackage{subcaption}
\usepackage{cleveref}
\usepackage{booktabs,siunitx,tabularx, stfloats}
\usepackage{flushend}
\usepackage{enumitem}
\usepackage[font=small,skip=5pt]{caption}

\crefname{section}{Section}{Sections}
\crefname{figure}{Figure}{Figures}
\crefname{table}{Table}{Tables}

\ninept

\usepackage{glossaries}
\newacronym{cnn}{CNN}{Convolutional Neural Network}
\newacronym{dl}{DL}{Deep Learning}
\newacronym{ai}{AI}{Artificial Intelligence}
\newacronym{dnn}{DNN}{Deep Neural Network}
\newacronym{nn}{NN}{Neural Network}
\newacronym{mmf}{MMF}{MultiMedia Forensics}
\newacronym{vc}{VC}{Voice Conversion}
\newacronym{tts}{TTS}{Text-to-Speech}
\newacronym{ann}{ANN}{Artificial Neural Network}
\newacronym{df}{DF}{Deepfake}
\newacronym{ml}{ML}{Machine Learning}
\newacronym{mfcc}{MFCC}{Mel Frequency Cepstral Coefficient}
\newacronym{stft}{STFT}{Short Time Fourier Transform}
\newacronym{cqcc}{CQCC}{Constant Q Cepstral Coefficient}
\newacronym{roc}{ROC}{Receiver Operating Characteristic}
\newacronym{auc}{AUC}{Area Under the Curve}
\newacronym{tp}{TP}{True Positive}
\newacronym{tn}{TN}{True Negative}
\newacronym{fp}{FP}{False Positive}
\newacronym{fn}{FN}{False Negative}
\newacronym{tpr}{TPR}{True Positive Rate}
\newacronym{fpr}{FPR}{False Positive Rate}
\newacronym{tnr}{TNR}{True Negative Rate}
\newacronym{vae}{VAE}{Variational Autoencoders}
\newacronym{gan}{GAN}{Generative Adversarial Networks}
\newacronym{stlt}{STLT}{Short-Term and Long-Term}
\newacronym{gru}{GRU}{Gated recurrent unit}
\newacronym{ser}{SER}{Speech Emotion Recognition}
\newacronym{dft}{DFT}{Discrete Fourier Transform}
\newacronym{mae}{MAE}{Mean Absolute Error}
\newacronym{svm}{SVM}{Support Vector Machines}
\newacronym{gnn}{GNN}{Graph Neural Network}
\newacronym{enf}{ENF}{Electric Network Frequency}
\newacronym{nlp}{NLP}{Natural Language Processing}
\newacronym{cm}{CM}{countermeasure}
\newacronym{mlp}{MLP}{Multi-Layer Perceptron}
\newacronym{ssl}{SSL}{Self-supervised Learning}
\newacronym{pf}{PF}{Partial Fake}
\newacronym{fft}{FFT}{Fast Fourier Transform}
\newacronym{eer}{EER}{Equal Error Rate}
\newacronym{ola}{OLA}{Overlap-and-Add}
\newacronym{lfcc}{LFCC}{Linear Frequency Cepstral Coefficient}
\newacronym{fc}{FC}{Fully Connected}
\newacronym{cqt}{CQT}{Constant-Q Transform}
\newacronym{vad}{VAD}{Voice Activity Detector}
\newacronym{add}{ADD}{Audio Deepfake Detection}

\title{
Analyzing the Impact of Splicing Artifacts in Partially Fake Speech Signals
}

\makeatletter
\def\name#1{\gdef\@name{#1\\}}
\makeatother
\name{{\em Viola Negroni, Davide Salvi, Paolo Bestagini, Stefano Tubaro}
}

\address{Dipartimento di Elettronica, Informazione e Bioingegneria  \\
Politecnico di Milano, 20133 Milan, Italy \\
{\small \tt \{viola.negroni, davide.salvi, paolo.bestagini, stefano.tubaro\}@polimi.it} }

\begin{document}

\maketitle


\begin{abstract}
Speech deepfake detection has recently gained significant attention within the multimedia forensics community.
Related issues have also been explored, such as the identification of partially fake signals, i.e., tracks that include both real and fake speech segments.
However, generating high-quality spliced audio is not as straightforward as it may appear.
Spliced signals are typically created through basic signal concatenation. This process could introduce noticeable artifacts that can make the generated data easier to detect.
We analyze spliced audio tracks resulting from signal concatenation, investigate their artifacts and assess whether such artifacts introduce any bias in existing datasets.
Our findings reveal that by analyzing splicing artifacts, we can achieve a detection EER of 6.16\% and 7.36\% on PartialSpoof and HAD datasets, respectively, without needing to train any detector.
These results underscore the complexities of generating reliable spliced audio data and lead to discussions that can help improve future research in this area.
\end{abstract}

\section{Introduction}
\label{sec:intro}

Automatic speaker verification and speech deepfake detection have gained significant importance in recent times. With the proliferation of sophisticated generation tools and the potential threats associated with their misuse, there is an urgent need to develop systems able to analyze the content they produce and prevent potential menaces and threats.
The multimedia forensics community has been actively working in this direction, leading to the development of various systems designed to analyze and detect generated speech content~\cite{cuccovillo2022open, bhagtani2022overview}. 

Alongside the primary task of discriminating real and fake speech signals, another problem has been gaining increasing attention lately: detecting partially fake signals.
This involves analyzing speech tracks and determining whether these are intact or spliced, i.e., generated by the concatenation of real and synthetic speech segments.
Such hybrid signals pose a significant threat, as they can deceive traditional speech deepfake detectors that are not specifically trained to address them. 
Additionally, they generate subtle deepfakes that need detailed temporal analysis to be spotted, as even the insertion of a brief spoofed segment can significantly alter the overall meaning of a speech.

Due to their creation process, partially fake audio can exhibit two different types of artifacts: \emph{intrinsic artifacts}, inherently produced by the different properties of the concatenated audio frames (e.g., real and fake signals, fakes generated by different models, etc.); and \emph{induced artifacts}, introduced by processing techniques applied during the splicing process (e.g., windowing effects, discontinuities at the joints, etc.).
Exploiting these artifacts, numerous systems have been proposed in the literature to address partially fake speech signals, each utilizing different approaches and architectures~\cite{zhang2023range, li2023convolutional, li2024audio}. Some methods rely on acoustic features extracted from the input audio~\cite{zhang2021initial, zhu2023local}, while others leverage pre-trained self-supervised networks as wav2vec~\cite{cai2023waveform}.
The extracted features are then processed using classification networks, such as 
CNNs~\cite{yi2021half}, 
LSTMs~\cite{zhang2021multi}, and transformers~\cite{yadav2024mdrt}.

In addition to the introduction of new detectors, novel datasets have also been released~\cite{zhang2022partialspoof, yi2021half}, and research challenges have been organized~\cite{yi2022add} to foster advancements in this field.
While these contributions are crucial in advancing the current state of the art, generating a dataset with no specific bias is far from being an easy task.
While intrinsic artifacts are almost unavoidable and tied to the properties of the original tracks, induced artifacts can be carefully mitigated by the forger using specific techniques at the synthesis stage (e.g., tapering, overlapping joined frames, etc.). However, induced artifacts may be challenging to conceal when creating large-scale datasets in bulk. This is problematic as these irregularities could make the forged signals easier to detect and potentially bias the detectors trained on them.

In this paper, we analyze \emph{induced artifacts} in spliced audio tracks and investigate their impact on state-of-the-art datasets and detectors.
We start by analyzing simple sinusoidal concatenation to investigate the artifacts introduced by this process and understand their underlying causes.
Then, we extend our study to more complex audio tracks, such as those used to create partially fake speech signals.
We evaluate two datasets in the field, PartialSpoof and HAD, assessing the presence of induced splicing artifacts and investigating whether these introduce bias.
Finally, we explore the potential issues this phenomenon may introduce on detectors, evaluating their performance in different setups.
Our findings highlight the complexity of the partially fake speech detection task and highlight the need for careful datasets and detectors development to ensure the practical effectiveness of such systems in real-world scenarios.

To summarise, the main contributions of this work are:
\begin{itemize}[leftmargin=*]
    \item We show that signal splicing techniques may introduce detectable artifacts, potentially creating biased tracks (Sec.~\ref{sec:analysis}).
    \item We show that state-of-the-art datasets are affected by this issue (Sec.~\ref{subsec:visual_artifacts}) and that a simple threshold-based detector can expose it (Sec.~\ref{subsec:dr}).
    \item We provide guidelines for processing data to create more realistic and challenging partially fake speech signals  (Sec.~\ref{sec:mitigation}).
    \item We benchmark different learning-based detectors to see if they are biased by the presence of induced artifacts in the training data (Sec.~\ref{sec:detectors}).
\end{itemize}
\section{Spectral Leakage and Splicing Artifacts}
\label{sec:analysis}

In this section we provide an intuitive understanding of the underlying mechanisms that raise \emph{induced artifacts} during the audio concatenation process. We start by introducing the fundamental concept of spectral leakage and progress to a simplistic example involving sinusoid concatenation. While this section provides an overview of the key concepts, it is by no means exhaustive in the explanation of the complex phenomena responsible for these artifacts.

\vspace{0.5em}
\noindent
\textbf{Spectral Leakage.}
Spectral leakage is a phenomenon observed in the frequency domain analysis when energy from a signal spreads across multiple frequency bins.
This effect occurs when a signal is framed with a window function under the incorrect assumption that it is periodic within the observation window.
This situation can happen in two scenarios: (a) the signal is periodic, but the length of the window is not an integer multiple of its period; (b) the signal is non-periodic.
The mismatch between the actual signal and the assumed periodic model results in a distortion of the frequency spectrum, causing energy to spread across frequency bins that would ideally have zero energy.

\begin{figure*}[b]
\centering
\includegraphics[width=.85\textwidth]{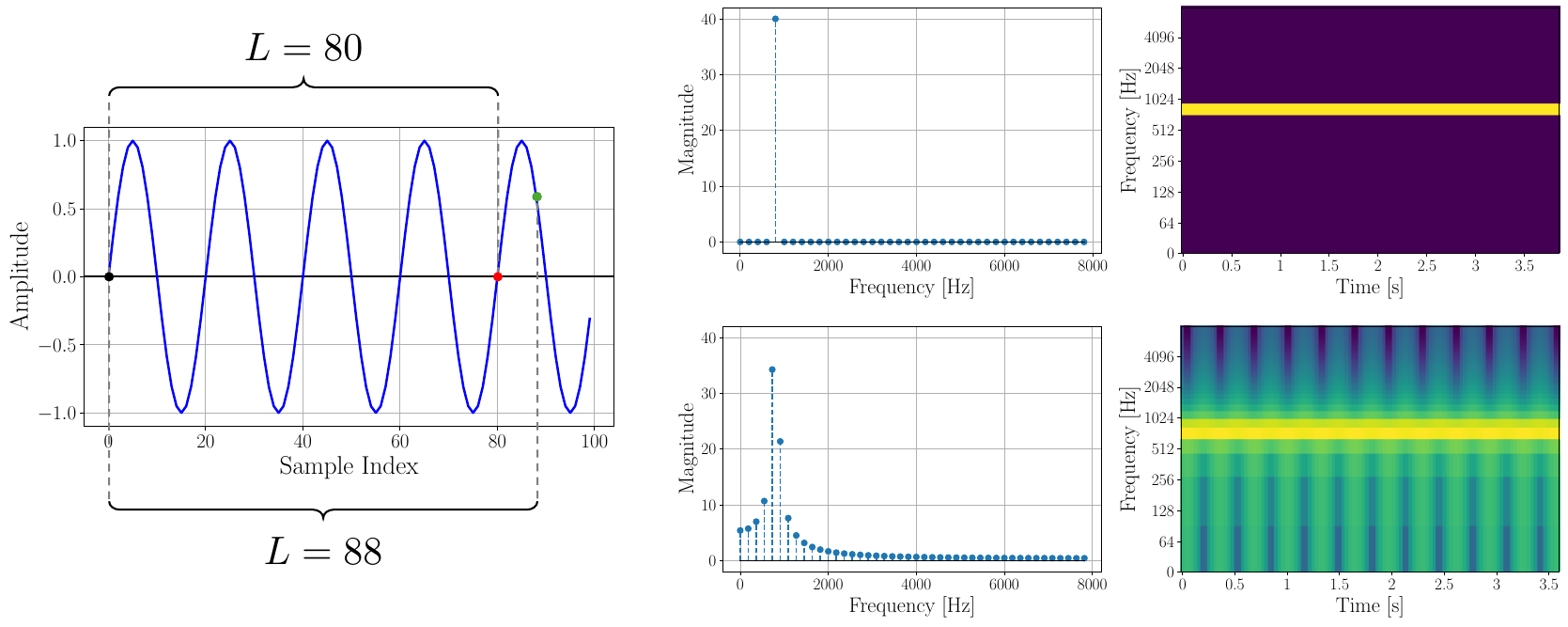} 
\caption{Frequency domain analysis of a sinusoid of frequency $f_0$ with two different analysis windows. $f_\text{s}$ = \SI{16}{\kilo\hertz}, $f_0$ = \SI{800}{\hertz}, period $T = 20$ samples. FFT up to Nyquist frequency and Spectrogram (dB).
Left: signal portion in the time domain. Center: DFT. Right: STFT magnitude. Top: STFT window of $L =80$ samples (multiple of $T$). Bottom: STFT window of $L =88$ samples (not a multiple of $T$), revealing spectral artifacts.}
\label{fig:pure_sin}
\end{figure*}

\iftrue
To illustrate the principle of spectral leakage, let us consider a discrete sinusoidal signal $\mathbf{x}$ and its \gls{dft} $\mathbf{X}$.
Ideally, the magnitude of $\mathbf{X}$ exhibits a single peak at the sinusoid's frequency. 
When $\mathbf{x}$ is framed with a window function $\mathbf{w}$, the resulting signal is 
    $\mathbf{x}_\text{w} = \mathbf{x} \cdot \mathbf{w}$,
where ``$\cdot$'' denotes sample-wise multiplication.
Let us denote with $\mathbf{W}$ the \gls{dft} of $\mathbf{w}$.
By definition, the \gls{dft} of $\mathbf{x}_\text{w}$ is given by
    $\mathbf{X}_\text{w} = \mathbf{X} * \mathbf{W}$,
where ``*'' indicates convolution.
The magnitude of $\mathbf{X}_\text{w}$ thus corresponds to the magnitude of $\mathbf{W}$ centered at the sinusoid's frequency.
If the window function $\mathbf{w}$ is rectangular, its \gls{dft} $\mathbf{W}$ corresponds to a $\text{sinc}$ function.
The $\text{sinc}$ function introduces sidelobes in the frequency domain of $\mathbf{X}_\text{w}$, causing energy to spread across adjacent frequency bins, a phenomenon known as spectral leakage.
In principle, we can minimize the spectral leakage by selecting a window whose length is a multiple of the signal's period.
This alignment ensures that the $\text{sinc}$ function is sampled at its zero crossing points, thereby reducing leakage.
However, when a signal is observed within a finite window with aperiodic boundary conditions, sidelobes still appear in the frequency spectrum.
This occurs because the $\text{sinc}$ function is no longer sampled precisely at its zero-crossings.
\fi
\cref{fig:pure_sin} illustrates this effect.
When the \gls{dft} is computed from an integer number of periods of the sinusoid, both the \gls{dft} magnitude and the spectrogram exhibit a distinct single peak.
In contrast, if the number of periods within the window is not an integer, a spectral leakage arises.


\vspace{0.5em}
\noindent
\textbf{Induced Splicing Artifact.}
To define a spliced recording, let us consider two discrete-time signals $\mathbf{x}_1$ and $\mathbf{x}_2$, with lengths $N_1$ and $N_2$, respectively. 
We define the spliced audio track $\mathbf{x}_\text{s}$ as the concatenation in time of $\mathbf{x}_1$ and $\mathbf{x}_2$, such as 
	$\mathbf{x}_\text{s} = [\mathbf{x}_1, \; \mathbf{x}_2]$.
The resulting length of $\mathbf{x}_\text{s}$ is $N = N_1 + N_2$, and the splicing point is at sample $N_1 + 1$, where the two signals are joined.
Speech forensic detectors commonly operate on spectrogram-based audio representations, which involve framing and windowing the speech under analysis.
We assert that in the case of partially fake signals, it is reasonable to assume that a forensic analyst will observe frames that span the splicing point. These frames are likely to exhibit induced splicing artifacts.

\begin{figure}[t]
    \centering
    \begin{subfigure}[b]{\columnwidth}
        \centering
        \includegraphics[width=\columnwidth]{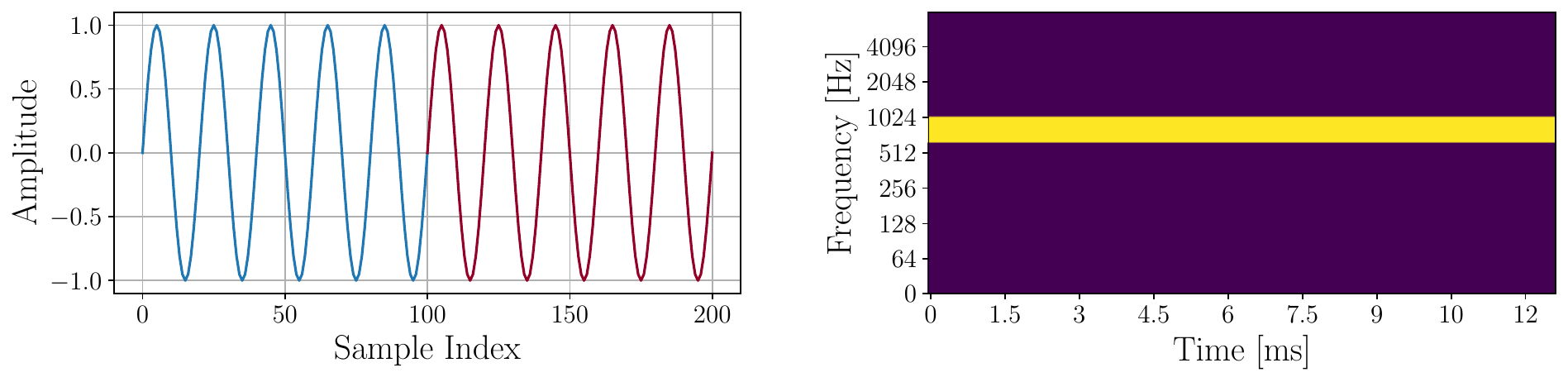} 
        \caption{$\mathbf{x}_1$ and $\mathbf{x}_2$ have same frequency, amplitude and phase.}
        \label{subfig:concat_ex0}
        \vspace{5pt}
    \end{subfigure}
    
    \begin{subfigure}[b]{\columnwidth}
        \centering
        \includegraphics[width=\columnwidth]{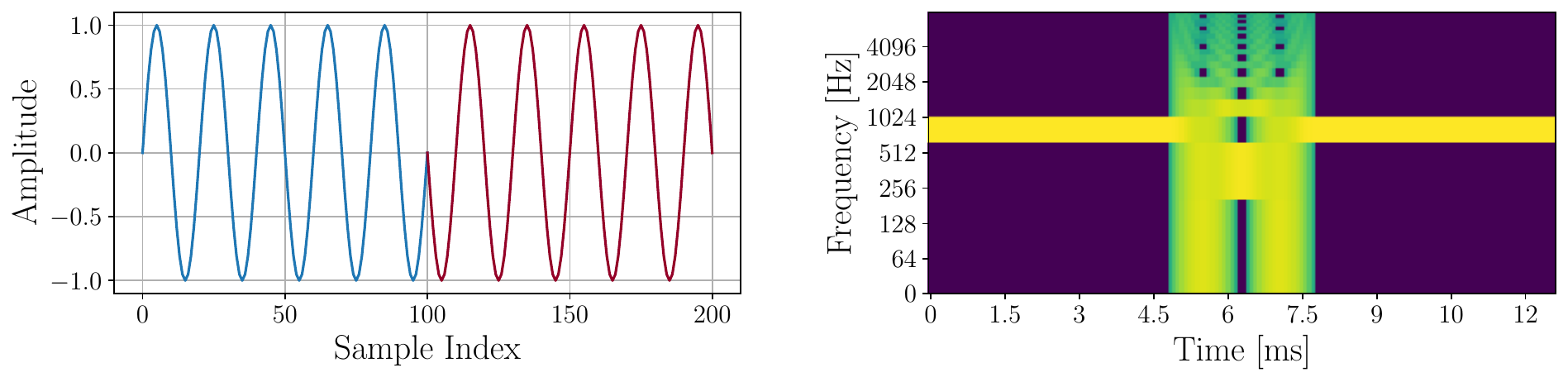} 
        \caption{$\mathbf{x}_1$ and $\mathbf{x}_2$ have same frequency and amplitude but different phase.}
        \label{subfig:concat_ex1}
        \vspace{5pt}
    \end{subfigure}
        
    \begin{subfigure}[b]{\columnwidth}
        \centering
        \includegraphics[width=\columnwidth]{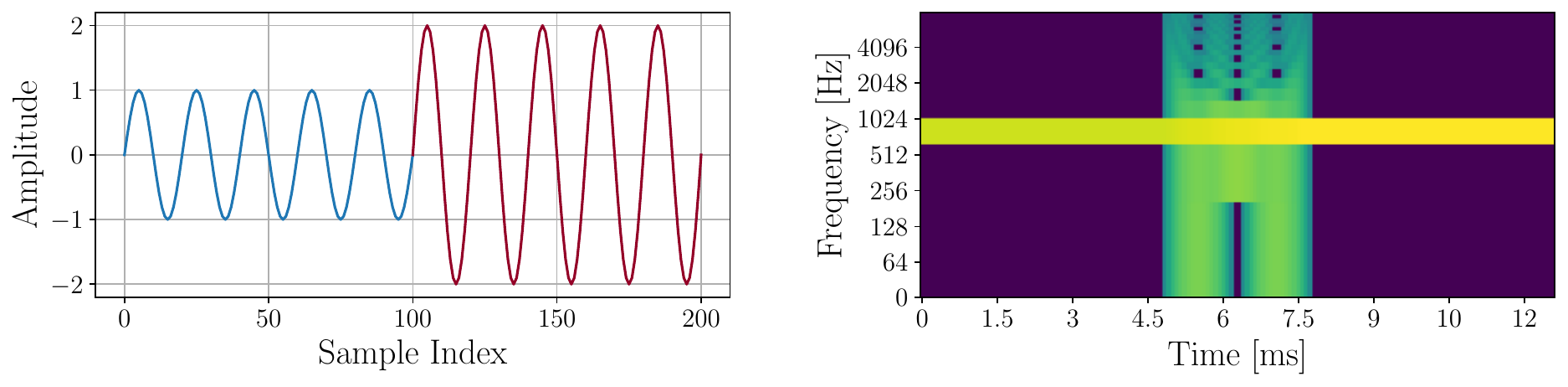} 
        \caption{$\mathbf{x}_1$ and $\mathbf{x}_2$ have same frequency and phase but different amplitude.}
        \label{subfig:concat_ex2}
    \end{subfigure}
    \caption{Frequency domain analysis of two concatenated sinusoids in different setups. Left: signal in the time domain. Right: STFT magnitude.
    }
    \label{fig:concat_examples}
\vspace{-15pt}
\end{figure}

Let us consider as an example an elementary spliced signal $\mathbf{x}_\text{s}$, which is created by concatenating a sinusoid $\mathbf{x}_1$ and a shifted version of itself $\mathbf{x}_2$ (e.g., by adding a phase shift). 
When framing this signal, even if we choose an analysis window whose length is equal to a multiple of the period of $\mathbf{x}_1$, the abrupt transition between the two sinusoids will cause leakage in the spectrum at the splicing point.
\cref{subfig:concat_ex1} shows an example of this phenomenon.
Let us consider another elementary signal $\mathbf{x}_\text{s}$ obtained by concatenating an integer amount of periods from a sinusoid $\mathbf{x}_1$ with a different amplitude version of the same sinusoid $\mathbf{x}_2$.
Even if we frame the signal with a window whose length is multiple of the sinusoid period, the amplitude change between the two segments will cause spectral leakage.
\cref{subfig:concat_ex2} shows an example of this phenomenon.

In this work, we define this specific type of spectral leakage as induced splicing artifact. Given that this leakage is evident even in simple sinusoidal concatenations, it is reasonable to expect that it will be magnified when dealing with complex signals, such as those used to create partially fake speech.
This means that if malicious users aim to create spliced tracks that do not contain any induced splicing artifact, they need to employ mitigation techniques at the synthesis stage.
It is important to note that although this study focuses on partially fake speech, these considerations apply to any signal regardless of whether the source samples are authentic or synthetically generated.

\section{Datasets Analysis}
\label{sec:datasets}

In this section, we explore the presence of induced artifacts in well-known datasets.
First, we introduce the datasets under consideration.
Then, we show that induced artifacts can be visually observed through data analysis.
Finally, we quantify the presence of induced artifacts in the datasets by distinguishing between spliced and non-spliced tracks using a straightforward method that exploits the artifacts.

\subsection{Considered Datasets}
We examine two state-of-the-art datasets designed for partially fake speech detection: PartialSpoof~\cite{zhang2022partialspoof} and HAD~\cite{yi2021half}.
To the best of our knowledge, these are the only two datasets available for partially fake speech detection at the time of writing.
Here, we provide a concise overview of both datasets, focusing on their construction policies. We refer the reader to the original papers for more detailed information.

\vspace{.5em}\noindent\textbf{PartialSpoof.}
PartialSpoof~\cite{zhang2022partialspoof} is an English speech database derived from the ASVspoof 2019 LA corpus~\cite{wang2020asvspoof}, which contains both real and partially fake speech signals.
It follows the same structure as the ASVspoof dataset and is divided into three partitions: training, development, and evaluation. 
The dataset includes synthetic speech generated by 17 distinct methods, varying between training-development and evaluation subsets.
The creation of the partially fake samples followed a rigorous 5-step procedure, summarized as follows: 
(1) Waveform amplitudes of real and fake utterances were normalized to \SI{-26}{dBov}.
(2) Variable-length candidate segments were chosen using three types of \gls{vad} methods, with the final selection based on majority voting. 
(3) Segments from real utterances were replaced with fake segments, and vice versa, using segments from different utterances by the same speaker.
Each segment was inserted only once per host utterance, and the inserted segments were of similar duration to the originals. 
Time-domain cross-correlation and the overlap-add method were employed for substitution and concatenation to minimize artifacts.
(4) After concatenation, each utterance was annotated with fine-grained segment labels and classified as real or partially fake, based on the presence or absence of synthetic patches. 
(5) Post-processing operations were performed in order to match the spoof class distribution of the ASVspoof 2019 LA database.

\vspace{.5em}\noindent\textbf{Half-truth Audio Detection.}
Half-truth Audio Detection (HAD)~\cite{yi2021half} is a Mandarin speech dataset and was the first corpus released containing partially fake speech tracks. 
It is based on AISHELL-3~\cite{shi21c_interspeech}, a multi-speaker speech dataset designed for training \gls{tts} models. 
While it was first released as an independent set, it later became part of the \gls{add} challenge dataset~\cite{yi2022add}. 
Partially fake speech tracks used in this corpus are created as follows.
First, the authors modified the transcripts of real speech samples by altering keywords to change the intended meaning.
Then, they trained a \gls{tts} model on AISHELL-3 to synthesize audio from the edited texts.  
Finally, they combined real and fake audio by substituting the edited keywords in the original speech with those from the synthetic recordings. The process included volume normalization and forced alignment to ensure precise replacement.

\subsection{Visual Artifacts Analysis}
\label{subsec:visual_artifacts}
\begin{figure*}
\centering
\includegraphics[width=\columnwidth]{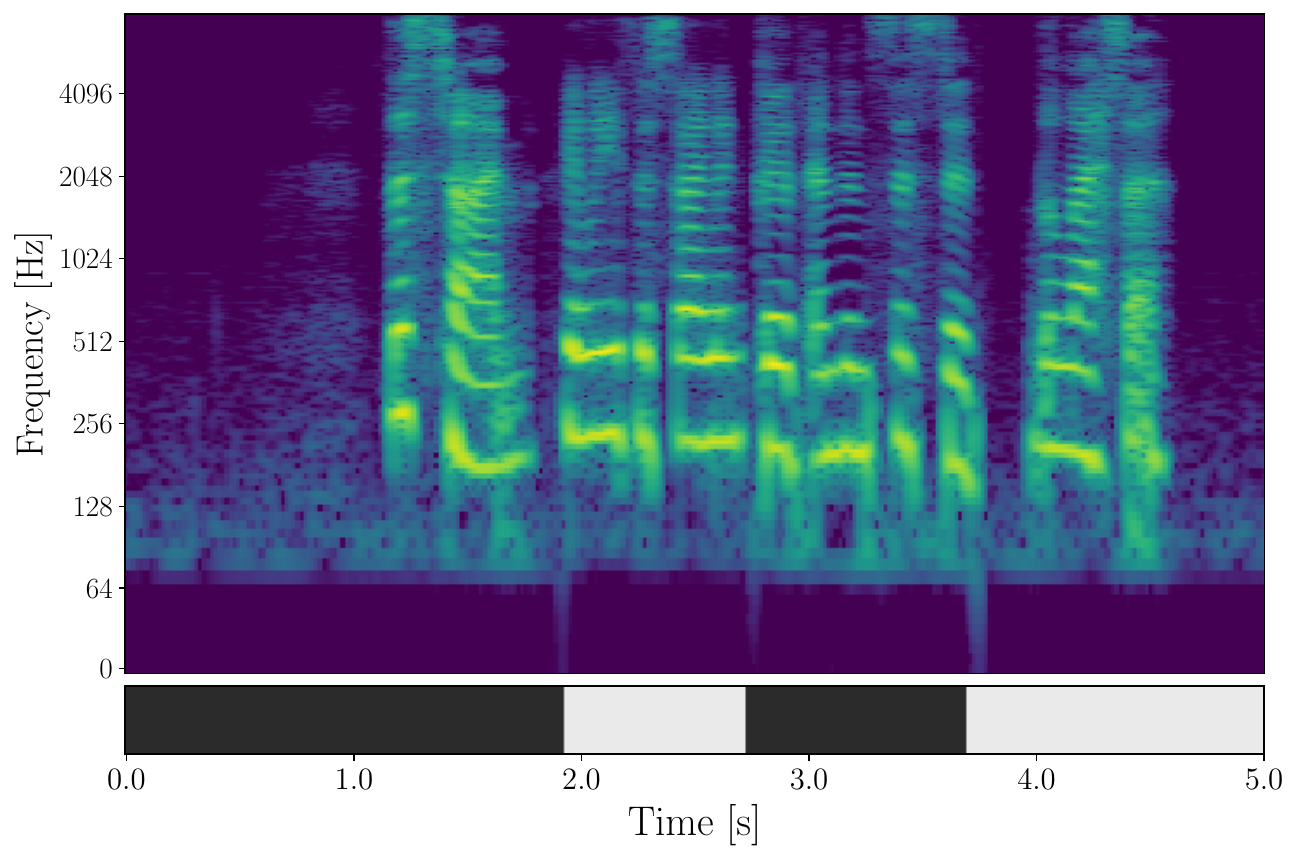} \quad
\includegraphics[width=\columnwidth]{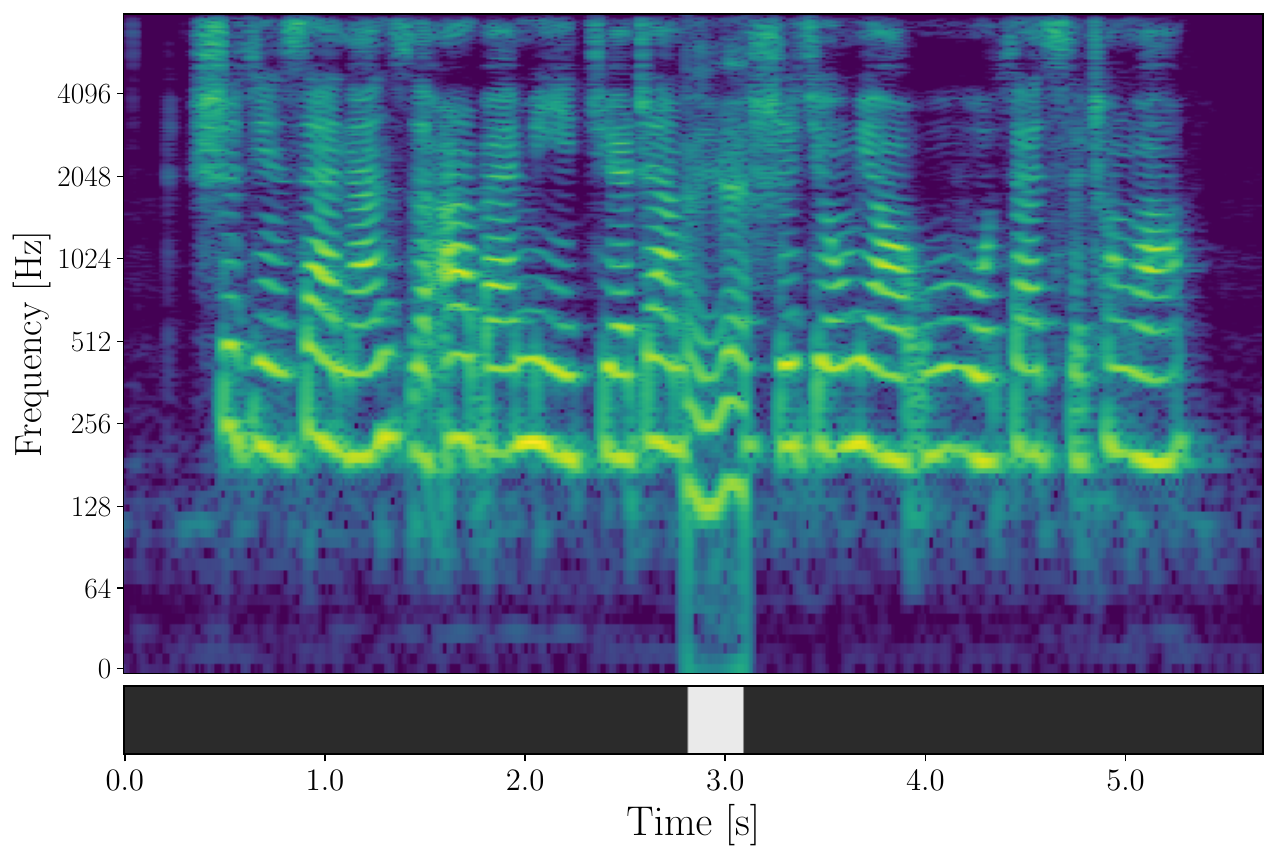} 
\caption{Frequency domain analysis of one example track per dataset: \textit{CON\_D\_0000001.wav} from PartialSpoof (left) and \textit{ADD2023\_T2\_D\_00000036.wav} from HAD (right).
STFT window size: 2048 samples, Hop size: 256 samples, no zero-padding. Top: full log-scaled spectrograms. Bottom: ground truth splicing timestamps, color changes indicate the splicing points.}
\label{fig:dataset_examples}
\end{figure*}

Let us consider the partially fake speech tracks from the two datasets presented above and investigate the presence of induced artifacts in their spliced tracks.
To this purpose, \cref{fig:dataset_examples} shows logarithmically scaled spectrograms of two partially fake speech samples, one from each dataset. 
Although the induced splicing artifacts at the concatenation points are generally inaudible, they can be easily exposed by a straightforward frequency analysis.



In the PartialSpoof track (left), the bias is particularly pronounced due to a distinct characteristic of the original ASVspoof 2019 dataset: all concatenated signals exhibit a nearly silent band at lower frequencies ($< \SI{80}{\hertz}$).
This characteristic makes it particularly challenging to hide any artifacts within this frequency range, as their presence becomes highly noticeable.
In contrast, the HAD track (right) does not show a significant silent band, yet induced artifacts are still visible.
Specifically, the leakages are pronounced in both the higher 
and lower frequency ranges,
where signals typically exhibit lower energy levels compared to the rest of the spectrum.
It is noteworthy how these artifacts persist despite the dataset's design, which includes various techniques to minimize the audibility of transitions between concatenated tracks.
This highlights the considerable difficulty in completely eliminating discontinuities caused by signal concatenations.

\subsection{Quantitative Artifacts Analysis}
\label{subsec:dr}

To quantify the presence of induced artifacts, we propose a simple method to discriminate between spliced and non-spliced tracks and evaluate its performance on the two considered datasets.
As discussed in the previous sections, induced splicing artifacts manifest as energy content spreading among spurious frequency components, which are visible in the frequency domain as ``streaks'' across the entire spectrum. 
Building on this insight, we examine specific frequency bands without speech content and measure their dynamic range.
Our hypothesis is that these bands, typically showing a shallow dynamic range due to the absence of speech, will exhibit an increased range in the presence of an induced artifact, i.e., a splicing point.

The proposed method to quantitative measure the presence of the aforementioned induced artifacts operates as follows.
Given an input speech signal $\mathbf{x}$, we compute its \gls{stft} and convert its amplitude in \si{\decibel}.
Let $\mathbf{X}_\text{dB}[k, m]$ be the STFT representation in \si{\decibel} of $\mathbf{x}$, where $k$ denotes the frequency bin and $m$ denotes the temporal frame index. 
We then select a subset of frequency bins and average their values, obtaining a vector $\mathbf{v}[m]$ with a unique value for each time index $m$, such as
\begin{equation}
	\mathbf{v}[m] = \frac{1}{|\mathcal{F}|} \sum_{k \in \mathcal{F}} \mathbf{X}_{\text{dB}}[k, m] ,
\end{equation} 
where $\mathcal{F}$ is the set of selected frequency bins.
Next, we compute the dynamic range $d$ of the vector $\mathbf{v}$ by calculating the difference between the maximum and minimum values of the averaged frequency bins.
By thresholding $d$, we discriminate between spliced and non-spliced signals.

\cref{tab:results} shows the results of the analysis using threshold-independent metrics, such as \gls{auc} and \gls{eer}.
The subsets $\mathcal{F}$ that we considered for each dataset were determined through experimental evaluation.
For the PartialSpoof dataset, we used the lowest \num{16} frequency bins of the \gls{stft} ($<\SI{60}{\hertz}$), while for HAD we used the highest \num{5} frequency bins ($>\SI{7960}{\hertz}$). 
We computed the \glspl{stft} considering a Hanning window with a length of \num{4096} (PS) and \num{2048} (HAD) samples, with no zero padding and a quarter of the window as hop length.  

The results we obtain are remarkable on both the considered datasets, with an average \gls{auc} of \num{98}\% across all the partitions of PartialSpoof and of \num{95}\% on those of the HAD dataset.
Additionally, we achieve competitive \glspl{eer} scores of 6.16\% and 7.36\% on the test partitions of the two corpora, respectively.
Our performance on the PartialSpoof evaluation partition outperforms that of the LCNN-based back-end system proposed in~\cite{zhang2021initial} (\gls{eer}=6.19\%)
and the SELCNN-based system of \cite{zhang2021multi} (\gls{eer}=6.33\%).
On the other hand, we are outperformed by the hybrid multiple-instance learning framework with local self-attention 
method proposed in~\cite{zhu2023local} (\gls{eer}=5.89\%), and by the wav2vec-based system presented in~\cite{zhang2022partialspoof} (\gls{eer}=0.49\%).
Regarding the HAD test partition, our \gls{eer} of 7.36\% again proves competitive. We outperform the GMM-based system proposed in~\cite{yi2021half} (\gls{eer}=12.67\%), while approaching the best-performing system of the ADD Challenge 2022,  based on wav2vec-XLS-R~\cite{lv2022fake} (\gls{eer}=4.8\%).
In analyzing these results, we want to highlight the simplicity of our proposed method compared to existing approaches.
Rather than utilizing cutting-edge models and extensive training processes, our approach requires minimal computational resources and relies on a straightforward analysis.
Also, we only examine a small portion of the input spectrogram, and we do not require any model training, while we effectively exploit a bias inherent in the considered data.

\begin{table}
\centering 
\caption{Detection of Partially Fake Speech Signals on PartialSpoof and HAD. $\uparrow$ means the higher the better. $\downarrow$ means the lower the better.}
\label{tab:results}
\resizebox{.95\columnwidth}{!}{\begin{tabular}{c|ccc|ccc}
\toprule
 & \multicolumn{3}{c|}{PartialSpoof} & \multicolumn{3}{c}{HAD} \\
 & Train         & Dev       & Eval        & Train       & Dev    &   Test  \\ \midrule 
\textbf{AUC (\%) $\uparrow$}    &   98.27   &   98.16   &   98.10    &    95.02   &   95.02  &  95.24  \\
\textbf{EER (\%) $\downarrow$}      &   6.16   &   5.57   &   6.16    &   5.30   &   7.36  &  7.36  \\ \bottomrule
\end{tabular}}
\vspace{-15pt}
\end{table}

\section{Mitigation Techniques}
\label{sec:mitigation}
Given the ease of detecting induced artifacts, we conducted experiments to determine if and to what extent it is possible to mitigate their presence.
The goal of artifact mitigation strategies is to conceal the artifacts by tampering with the frequency bands where they occur.
To do so, we built a small dataset of spliced audio by concatenating randomly selected real and fake segments, ensuring each generated track had only one splicing point.
Then, we tested various mitigation techniques, including lossy compression, cross-fading, \gls{ola}, and linear predictive coding, and analyzed the obtained tracks to determine the effectiveness of each.
In the following, we focus on the approach that proved most effective in concealing artifacts: \gls{ola}.
To create the dataset, we selected tracks from the train set of ASVspoof 2019. We did so to ensure the most challenging scenario possible, given that these tracks contain a completely silent band, making it particularly difficult to conceal induced artifacts.
We generated the spliced tracks with the following approach:
\begin{enumerate}[leftmargin=*,noitemsep,nolistsep]
    \item We select one real and one fake track from the same speaker and then use a \gls{vad} to find the longest silent region within each track (excluding leading and trailing silences). The concatenation is performed within these silent regions, as they exhibit lower energy levels compared to the rest of the track.
    \item We randomly choose either the real or the fake track and extract the segment from the beginning to the end of its longest silence region. For the other track, we retain the portion from the beginning of its longest silence to the end.
    \item We apply \gls{ola} on the two segments using a Hanning window, applying half of the window function to the end of the first segment and the other half at the beginning of the second segment.
\end{enumerate}
We experimented with five different lengths of \gls{ola} windows, creating \num{1200} spliced tracks for each configuration.
We evaluated our detection method on these tracks and other \num{1200}  real tracks randomly selected from the remaining dataset. 
The experimental setup for the dynamic range analysis is the same as outlined in \cref{subsec:dr} for the PartialSpoof dataset.
Results for each \gls{ola} window are reported in the \emph{Clean} column of \cref{tab:ola}.

Considering an \gls{ola} window of \num{256} samples, the \gls{auc} value aligns with that obtained on the PartialSpoof dataset (AUC=98\%). 
While increasing the window size helps to mitigate the presence of the artifacts, the minimum achieved \gls{auc} is still a substantial \num{88.99}\%.
Therefore, we repeated the experiment by injecting different levels of white noise ($\text{SNR}_{\text{dB}} 60$, $\text{SNR}_{\text{dB}} 50$, $\text{SNR}_{\text{dB}} 46$, $\text{SNR}_{\text{dB}} 40$) into the signals. 
We applied a low-pass filter of order 7 with a cut-off frequency of \SI{80}{\hertz} on the noise so that it applied only to the lower frequencies of the signals and remained inaudible. 
As shown from columns 3 to 6 of \cref{tab:ola}, effectively hiding the artifacts requires both a substantial level of noise, such as $\text{SNR}_{\text{dB}} 46$ or $\text{SNR}_{\text{dB}} 40$, and an \gls{ola} window of at least \num{1024} samples. 
With the shortest window of \num{256} samples and maximum noise $\text{SNR}_{\text{dB}} 40$, the \gls{auc} value still stands at \num{70.51}\%.
As a final experiment, we applied a high-pass filter to the partially fake tracks we generated, considering a filter of order \num{8} with a cut-off at \SI{100}{\hertz}. 
\gls{auc} values are shown in the last column of \cref{tab:ola}. 
As widely expected, this brute-force method makes the bias introduced by the artifacts undetectable, rendering the dynamic range analysis ineffective regardless of the chosen \gls{ola} window.

\section{Detectors Analysis}
\label{sec:detectors}



Given that a straightforward threshold-based detector performs effectively on partially fake speech tracks due to the presence of induced artifacts, we now want to investigate whether these leakages might introduce bias into more advanced detectors trained and tested on these datasets.
It has already been shown that partially fake speech detectors focus on the splicing points to perform their prediction~\cite{liu2024neural}. However, we aim to analyze whether this focus is due to the presence of critical information at these points or merely because of the induced splicing artifacts.
If the focus is on critical information, this could lead to valuable insights. On the other hand, if it is due to artifacts, it would indicate a bias in the trained detectors.
To verify this aspect, we evaluate four state-of-the-art speech deepfake detectors, each utilizing different input features.
We train the models for the partially fake speech detection task, focusing on the so-called utterance-level classification, i.e., discriminating between authentic and partially fake tracks.

\begin{table}
\centering 
\large
\caption{Percentage AUC values from dynamic range analysis on the custom dataset with different OLA window sizes and setups: no post-processing (Clean), white noise injection ($\text{SNR}_{\text{dB}} \text{XX}$), and high-pass filtering (High-Pass).}
\label{tab:ola}
\resizebox{\columnwidth}{!}
{\begin{tabular}{c|c|cccc|c}
\toprule
{OLA Win.} & Clean & $\text{SNR}_{\text{dB}} 60$ & $\text{SNR}_{\text{dB}} 50$ & $\text{SNR}_{\text{dB}} 46$ & $\text{SNR}_{\text{dB}} 40$ & High-Pass  \\ \midrule 
256   &  98.04 & 96.89 & 87.45 & 81.74 & 70.51 & 56.03  \\
512   &  96.15 & 93.86 & 76.82 & 70.85 & 62.10 & 55.29  \\
1024  &  91.93 & 86.73 & 62.88 & 60.30 & 57.41 & 55.76 \\
2048  &  88.31 & 81.75 & 59.55 & 57.21 & 55.84 & 55.66 \\
4096  &  88.99 & 81.44 & 59.44 & 57.86 & 54.29 & 56.03 \\ \bottomrule
\end{tabular}}
\vspace{-15pt}
\end{table}

The considered models include RawNet2~\cite{tak2021end} which processes raw waveforms, a LCNN model~\cite{wu2018light} which receives \glspl{lfcc} as input, a SENet34~\cite{hu2018squeeze} fed with spectrograms and MCG-Res2Net50~\cite{li2021channel} trained on \gls{cqt} representations of the input audio.
We trained all the models for \num{100} epochs, using Cross-Entropy as loss function Adam as optimizer.
Validation loss was monitored throughout training, with early stopping set to \num{20} epochs and a learning rate of \num{e-4} appropriately reduced on plateaux.
Batch sizes were \num{128}, \num{246}, \num{48}, and \num{64} samples for RawNet2, LCNN, SENet34, and MCG-Res2Net50, respectively, and each of these was balanced to contain the same number of authentic and spliced samples.
We trained and validated the detectors on the training and development partitions of PartialSpoof, respectively. 
We decided to focus this analysis on PartialSpoof only, as it is the most widely employed dataset for the task at hand.
Then, we tested the models on two different versions of the evaluation partition: the original data and a version where we applied the same high-pass filtering described in \cref{sec:mitigation} to remove the induced artifacts from the spliced tracks.
The goal of this experiment was to determine whether the detectors rely on splicing artifacts for their predictions or if they focus more on the actual signal content.

\begin{figure}
    \centering
        \includegraphics[width=.75\columnwidth]{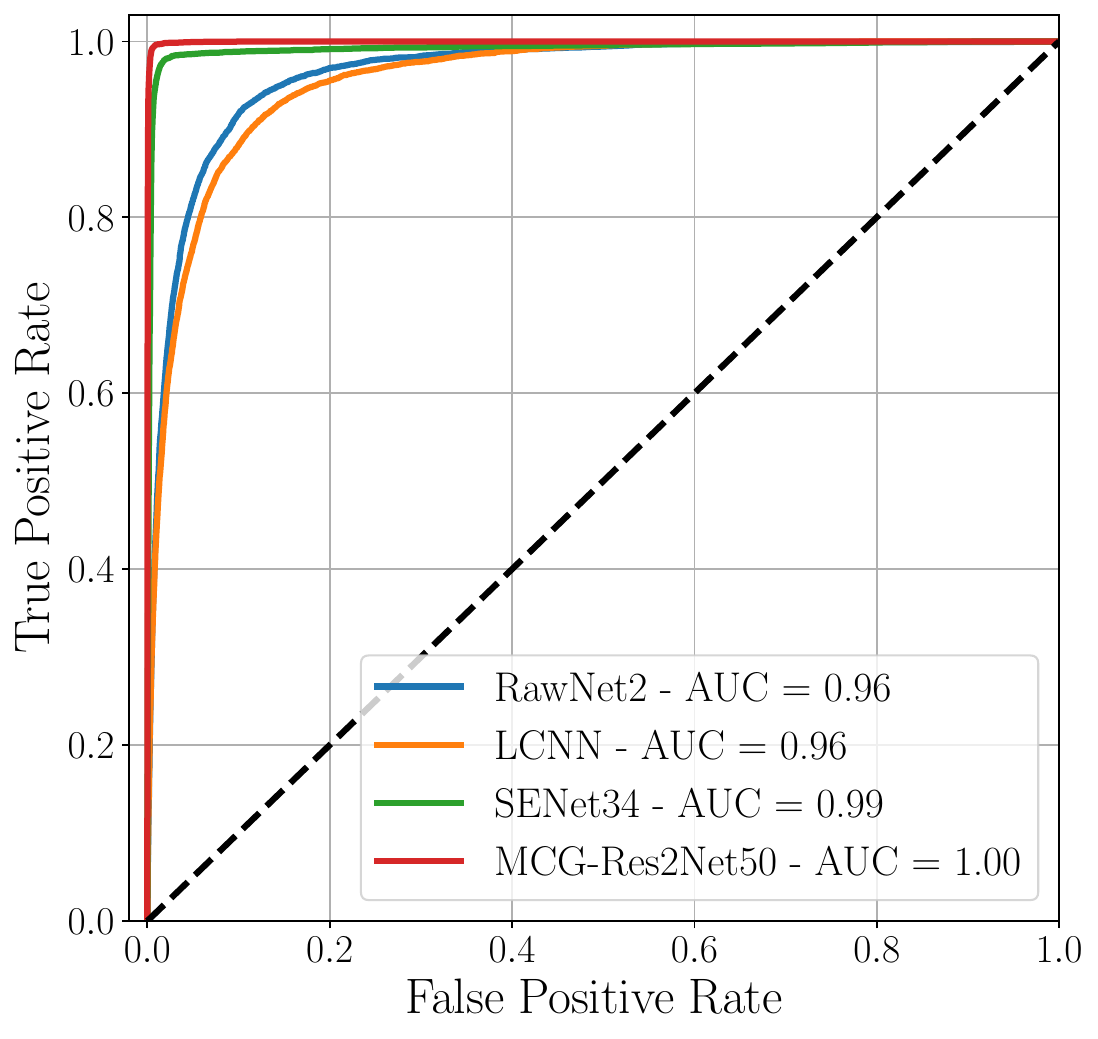} 
        \caption{ROC and AUC values of the detectors trained and tested on the original PartialSpoof dataset.}
        \label{fig:det_exp_1}
\vspace{-15pt}
\end{figure}

\begin{figure}
    \centering
        \includegraphics[width=.75\columnwidth]{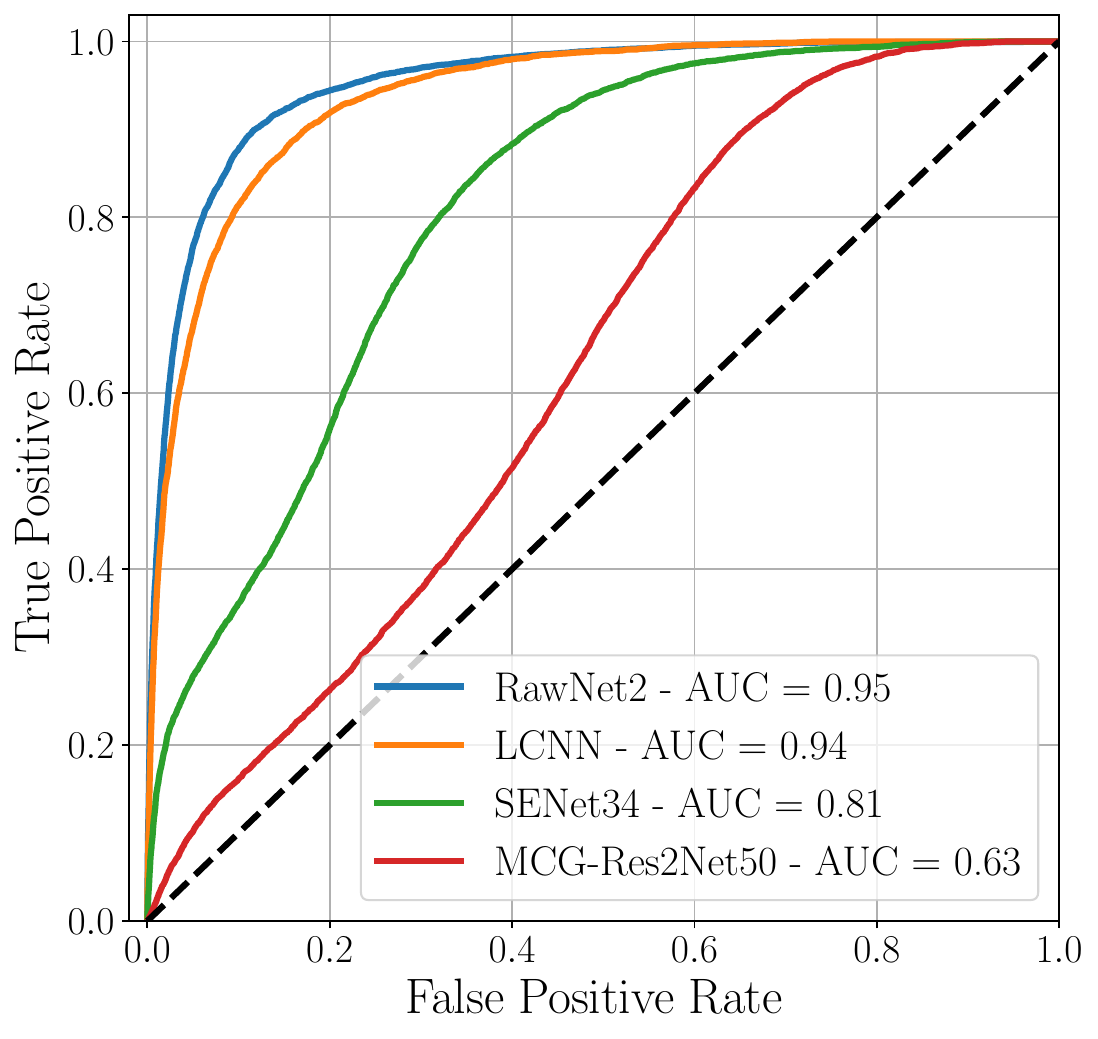} 
        \caption{ROC and AUC values of the detectors trained on the original PartialSpoof dataset and tested on its high-pass filtered version.}
        \label{fig:det_exp_2}
\vspace{-15pt}
\end{figure}

\cref{fig:det_exp_1} and \cref{fig:det_exp_2} show the results of this analysis.
The detectors trained using raw waveforms and \glspl{lfcc} as input, namely RawNet2 and the LCNN, prove robust when induced splicing artifacts are filtered out from the evaluation data, with an \gls{auc} value that is almost not affected.
In contrast, the performances of the SENet34 and MCG-Res2Net50 drop significantly when the artifacts are removed from the test set: the \gls{eer} of the SENet34 increases from \num{2}\% to \num{27.2}\%, and that of MCG-Res2Net50 goes from a remarkable \num{0.6}\% to \num{43.2}\%, that is close to random guessing. 
This outcome suggests that the SENet34 and MCG-Res2Net50 have been significantly biased by the presence of induced artifacts in the training data.
It is no coincidence that the models showing the most noticeable performance deterioration are those that rely on input features that explicitly consider the frequency domain, i.e., spectrograms and \glspl{cqt}. 
These models, which analyze detailed frequency information, are more susceptible to overfitting on spurious frequency components, like the induced artifacts, affecting their robustness when such leakages are absent.

\begin{figure}
    \centering
        \includegraphics[width=.75\columnwidth]{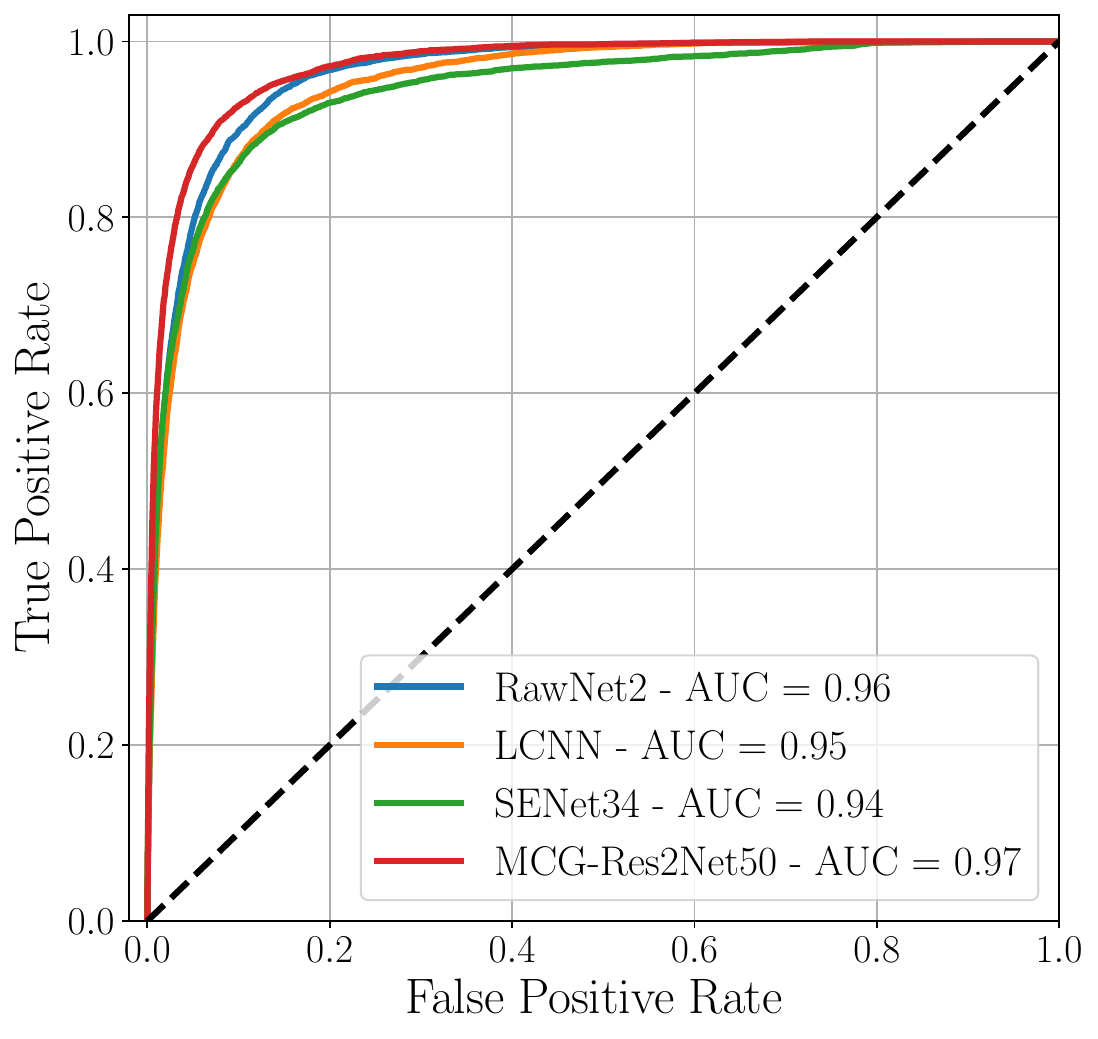} 
        \caption{ROC curves and AUC values for the detectors trained and tested on the high-pass filtered version of PartialSpoof.}
        \label{fig:detectors_exp_3}
\vspace{-15pt}
\end{figure}

As a final experiment, we investigated to what extent the spectral leakage bias boosted the performance of the detectors.
We re-trained the detectors on PartialSpoof, this time applying the high-pass filter to both the training and test sets.
The seed and training setup for all models remained unchanged. 
\cref{fig:detectors_exp_3} shows the results of this analysis. 
After filtering, the performance of SENet34 and MCG-ResNet50 improved and is now more comparable to that of RawNet2 and the LCNN, suggesting that they are no longer overfitting to any bias.
These findings highlight the importance of proper training to ensure detectors avoid bias. This is a crucial aspect for applying the developed models to real-world scenarios where such artifacts may be minimized, and detection cannot rely on their presence.

\section{Discussion}
\label{sec:discussion}

Induced artifacts resulting from signal concatenation are challenging to avoid when generating a spliced signal. Using a sufficiently short \gls{stft} analysis window, frequency smearing at the concatenation point will occur unless a perfectly continuous signal is built in both terms of magnitude and phase components (\cref{subfig:concat_ex0}).
Therefore, creating a high-quality spliced track from concatenation is not as straightforward as it may seem since even minor discontinuities in the resulting signal can introduce spurious frequency components in its spectrum.
The presence of such induced artifacts can make a spliced track extremely easy to detect.
The difficulty in avoiding this phenomenon has both positive and negative implications. On the positive side, it suggests that detecting splicing attacks by a malicious user may be easier than expected, especially if the attacker lacks expertise in signal processing.
On the negative side, the presence of induced artifacts in partially fake tracks can bias state-of-the-art datasets and lead detectors to overfit these artifacts, particularly those that heavily rely on frequency domain inputs.
Avoiding this bias is crucial, as there is a valid concern that these detectors may perform poorly when evaluated on spliced data where induced artifacts have been successfully concealed.
One key aspect to avoid the detectability of induced artifacts is the choice of source data for constructing spliced samples. Selecting source tracks that are rich in content across all frequency components can effectively conceal artifacts, as the dense spectrum of these signals is less impacted by the introduction of artifacts. To prove this point, we repeated the experiment from \cref{sec:mitigation} using tracks from the training set of the newly released ASVspoof5 dataset~\cite{Wang2024_ASVspoof5} as source data, following the same setup as before. We achieved an average \gls{auc} of \num{57}\% for all the \gls{ola} windows without applying any noise or filtering.
This result is significantly better compared to those obtained in \cref{sec:detectors} (average \gls{auc}=\num{92.6}\%) and highlights the critical role of source data in mitigating the presence of induced artifacts in spliced tracks, which is pivotal to ensure the development of synthetic speech detectors that do not overfit them.
\section{Conclusions}
\label{sec:conclusions}

In this work, we explored the nature of spectral artifacts resulting from the concatenation of different signals.
We evaluated the presence of these artifacts in state-of-the-art datasets for partially fake speech detection and their effects on synthetic speech detectors trained for this task.
Our findings indicate that both the partially fake tracks of PartialSpoof and the HAD dataset are significantly affected by this phenomenon, and it is possible to discriminate between their real and spliced tracks with a straightforward dynamic range analysis without the need to train any detector.
Also, we show that such induced splicing artifacts can create a bias on which synthetic speech detectors might overfit, especially those trained on frequency domain input features. 
We investigated several mitigation methods and found that, while completely eliminating these artifacts is almost impossible, applying basic signal processing smoothing techniques and, most importantly, carefully selecting the source data used to generate the spliced tracks can effectively address the issue.

\section{Acknowledgements}


This research is sponsored by the Defense Advanced
Research Projects Agency (DARPA) and the Air Force Research Laboratory
(AFRL) under agreement number FA8750-20-2-1004.
This work
was partially supported by the European Union under the Italian National
Recovery and Resilience Plan (NRRP) of NextGenerationEU
(PE00000001 - program ``RESTART'' and PE00000014 - program ``SERICS'') and 
by the ``FOSTERER'' project, funded by the Italian Ministry of University and Research within PRIN 2022 program.

\newpage

\bibliographystyle{IEEEbib}
\bibliography{references}

\begin{thebibliography}{10}

\bibitem{cuccovillo2022open}
Luca Cuccovillo, Christoforos Papastergiopoulos, Anastasios Vafeiadis, Artem Yaroshchuk, Patrick Aichroth, Konstantinos Votis, and Dimitrios Tzovaras,
\newblock ``Open challenges in synthetic speech detection,''
\newblock in {\em IEEE International Workshop on Information Forensics and Security (WIFS)}, 2022.

\bibitem{bhagtani2022overview}
Kratika Bhagtani, Amit Kumar~Singh Yadav, Emily~R. Bartusiak, Ziyue Xiang, Ruiting Shao, Sriram Baireddy, and Edward~J. Delp,
\newblock ``{An Overview of Recent Work in Multimedia Forensics},''
\newblock in {\em IEEE Conference on Multimedia Information Processing and Retrieval}, 2022.

\bibitem{zhang2023range}
Lin Zhang, Xin Wang, Erica Cooper, Nicholas Evans, and Junichi Yamagishi,
\newblock ``Range-based equal error rate for spoof localization,''
\newblock in {\em Interspeech}. ISCA, 2023.

\bibitem{li2023convolutional}
Kang Li, Xiao-Min Zeng, Jian-Tao Zhang, and Yan Song,
\newblock ``Convolutional recurrent neural network and multitask learning for manipulation region location.,''
\newblock in {\em DADA@ IJCAI}, 2023, pp. 18--22.

\bibitem{li2024audio}
Menglu Li, Yasaman Ahmadiadli, and Xiao-Ping Zhang,
\newblock ``Audio anti-spoofing detection: A survey,''
\newblock {\em arXiv preprint arXiv:2404.13914}, 2024.

\bibitem{zhang2021initial}
Lin Zhang, Xin Wang, Erica Cooper, Junichi Yamagishi, Jose Patino, and Nicholas Evans,
\newblock ``An initial investigation for detecting partially spoofed audio,''
\newblock in {\em Interspeech}. ISCA, 2021.

\bibitem{zhu2023local}
Yupeng Zhu, Yanxiang Chen, Zuxing Zhao, Xueliang Liu, and Jinlin Guo,
\newblock ``Local self-attention-based hybrid multiple instance learning for partial spoof speech detection,''
\newblock {\em ACM Transactions on Intelligent Systems and Technology}, vol. 14, no. 5, pp. 1--18, 2023.

\bibitem{cai2023waveform}
Zexin Cai, Weiqing Wang, and Ming Li,
\newblock ``Waveform boundary detection for partially spoofed audio,''
\newblock in {\em ICASSP 2023-2023 IEEE International Conference on Acoustics, Speech and Signal Processing (ICASSP)}. IEEE, 2023, pp. 1--5.

\bibitem{yi2021half}
Jiangyan Yi, Ye~Bai, Jianhua Tao, Haoxin Ma, Zhengkun Tian, Chenglong Wang, Tao Wang, and Ruibo Fu,
\newblock ``Half-truth: A partially fake audio detection dataset,''
\newblock in {\em Interspeech}. ISCA, 2021.

\bibitem{zhang2021multi}
Lin Zhang, Xin Wang, Erica Cooper, and Junichi Yamagishi,
\newblock ``Multi-task learning in utterance-level and segmental-level spoof detection,''
\newblock in {\em Interspeech}. ISCA, 2021.

\bibitem{yadav2024mdrt}
Amit Kumar~Singh Yadav, Kratika Bhagtani, Sriram Baireddy, Paolo Bestagini, Stefano Tubaro, and Edward~J Delp,
\newblock ``{MDRT: Multi-domain synthetic speech localization},''
\newblock in {\em IEEE International Conference on Acoustics, Speech and Signal Processing (ICASSP)}, 2024.

\bibitem{zhang2022partialspoof}
Lin Zhang, Xin Wang, Erica Cooper, Nicholas Evans, and Junichi Yamagishi,
\newblock ``The partialspoof database and countermeasures for the detection of short fake speech segments embedded in an utterance,''
\newblock {\em IEEE/ACM Transactions on Audio, Speech, and Language Processing}, vol. 31, pp. 813--825, 2022.

\bibitem{yi2022add}
Jiangyan Yi, Ruibo Fu, Jianhua Tao, Shuai Nie, Haoxin Ma, Chenglong Wang, Tao Wang, Zhengkun Tian, Ye~Bai, Cunhang Fan, et~al.,
\newblock ``{ADD 2022: the first audio deep synthesis detection challenge},''
\newblock in {\em IEEE International Conference on Acoustics, Speech and Signal Processing (ICASSP)}, 2022.

\bibitem{wang2020asvspoof}
Xin Wang, Junichi Yamagishi, Massimiliano Todisco, H{\'e}ctor Delgado, Andreas Nautsch, Nicholas Evans, Md~Sahidullah, Ville Vestman, Tomi Kinnunen, Kong~Aik Lee, et~al.,
\newblock ``Asvspoof 2019: A large-scale public database of synthesized, converted and replayed speech,''
\newblock {\em Computer Speech \& Language}, vol. 64, pp. 101--114, 2020.

\bibitem{shi21c_interspeech}
Yao Shi, Hui Bu, Xin Xu, Shaoji Zhang, and Ming Li,
\newblock ``{AISHELL-3: A Multi-Speaker Mandarin TTS Corpus},''
\newblock in {\em Interspeech}. ISCA, 2021.

\bibitem{lv2022fake}
Zhiqiang Lv, Shanshan Zhang, Kai Tang, and Pengfei Hu,
\newblock ``{Fake audio detection based on unsupervised pretraining models},''
\newblock in {\em IEEE International Conference on Acoustics, Speech and Signal Processing (ICASSP)}, 2022, pp. 9231--9235.

\bibitem{liu2024neural}
Tianchi Liu, Lin Zhang, Rohan~Kumar Das, Yi~Ma, Ruijie Tao, and Haizhou Li,
\newblock ``How do neural spoofing countermeasures detect partially spoofed audio?,''
\newblock in {\em Interspeech}. ISCA, 2024.

\bibitem{tak2021end}
Hemlata Tak, Jose Patino, Massimiliano Todisco, Andreas Nautsch, Nicholas Evans, and Anthony Larcher,
\newblock ``End-to-end anti-spoofing with rawnet2,''
\newblock in {\em IEEE International Conference on Acoustics, Speech and Signal Processing (ICASSP)}, 2021.

\bibitem{wu2018light}
Xiang Wu, Ran He, Zhenan Sun, and Tieniu Tan,
\newblock ``{A light CNN for deep face representation with noisy labels},''
\newblock {\em {IEEE Transactions on Information Forensics and Security}}, vol. 13, no. 11, pp. 2884--2896, 2018.

\bibitem{hu2018squeeze}
Jie Hu, Li~Shen, and Gang Sun,
\newblock ``Squeeze-and-excitation networks,''
\newblock in {\em IEEE Conference on Computer Vision and Pattern Recognition (CVPR)}, 2018.

\bibitem{li2021channel}
Xu~Li, Xixin Wu, Hui Lu, Xunying Liu, and Helen Meng,
\newblock ``Channel-wise gated res2net: Towards robust detection of synthetic speech attacks,''
\newblock in {\em Interspeech}. ISCA, 2021.

\bibitem{Wang2024_ASVspoof5}
Xin Wang, H{\'e}ctor Delgado, Hemlata Tak, Jee-weon Jung, Hye-jin Shim, Massimiliano Todisco, Ivan Kukanov, Xuechen Liu, Md~Sahidullah, Tomi Kinnunen, Nicholas Evans, Kong~Aik Lee, and Junichi Yamagishi,
\newblock ``{ASVspoof 5}: {Crowdsourced} speech data, deepfakes, and adversarial attacks at scale,''
\newblock in {\em ASVspoof Workshop 2024 (accepted)}, 2024.

\end{thebibliography}

\end{document}